\documentclass[a4paper,12pt]{article}
\usepackage{amsmath,amsfonts,amssymb}
\usepackage{relsize}

\begin{document}

\begin{center}
{\LARGE\bf A novel multiquark approach to hadron resonances}
\end{center}

\bigskip

\begin{center}
{\large S. S. Afonin
%\footnote{Email: \texttt{s.afonin@spbu.ru}.}
%\setcounter{footnote}{0}
}
\end{center}

\begin{center}
{\it Saint Petersburg State University, 7/9 Universitetskaya nab.,
St.Petersburg, 199034, Russia\\
E-mail:} \texttt{s.afonin@spbu.ru}
\end{center}

\bigskip

\begin{abstract}
A new approach to description of hadron spectroscopy is proposed.
By assumption, the form of spectrum is dictated by the trace of
energy momentum tensor in QCD. This provides the relativistic and
renormalization invariance of hadron masses. The constructed
scheme is applied to the light mesons for which two complementary
interpretations are worked out. The first one represents an
"atomic" structure of resonances above 1~GeV in which the quanta
of non-perturbative gluon contributions are quantified via an
effective formation of quasiparticles representing likely gluon
analogues of positronium. This picture allows to build a "periodic
table of the hadronic elements", i.e. to classify hadrons, in some
sense, in analogy with Mendeleev table in chemistry. The
classification does not require introduction of the orbital
angular momentum associated with hadron constituents. The Regge
and radial trajectories emerge in a natural way. The second
interpretation is based on a "collisional" nature of some (or
many) hadrons, specifically of scalar resonances below 1~GeV. Here
the role of quasiparticles is played by another hadron. This
picture in particular leads to a simple explanation of the
puzzling scalar sector below 1~GeV with correct predictions for
masses and dominant decay modes.
\end{abstract}

\newpage

\section{Introduction}

The strong interactions between light $u$ and $d$ quarks are known
to generate almost 99\% of mass of the visible universe. In spite
of many successes of Quantum Chromodynamics, the mechanism of
hadron mass generation remains unclear. The problem is the most
pronounced in the sector of light quarks where the underlying
physics is highly non-perturbative and ultrarelativistic. The
production of a large number of hadron resonances observed in the
hadron collisions and leptonic processes~\cite{pdg} provides a
striking manifestation of this physics. It is widely believed that
the mass generation mechanism is enciphered in the spectroscopy of
resonances. A wish to understand the rich hadron spectroscopy has
driven an active model building since the 1960s. And although a
great deal of efforts was invested in this enterprise in the last
fifty years, a generally accepted analytical approach that would
describe systematically the whole hadron spectrum has still not
been elaborated. Such a situation could mean that some basic
concepts which are being used for interpretation of hadron
resonances and for model building may happen to be somewhat
incomplete or even partly wrong. The proliferating observations of
unconventional resonances in the heavy quark sector strengthens
this impression~\cite{chen}. It is not unlikely that for further
advancing in the field of hadron spectroscopy one needs
essentially new ideas and approaches.

Our research was inspired by the following problems. The standard
interpretation of observable hadrons includes the non-relativistic
notion of orbital angular momentum $L$ associated with quarks. In
particular, $L$ dictates the spatial and charge parities for the
quark-antiquark systems, $P=(-1)^{L+1}$ and $C=(-1)^{L+S}$, where
$S$ is the quark spin. This picture does not explain the scalar
mesons below 1~GeV for which a tetraquark structure is often
assumed~\cite{pelaez}. A question arises why do not we see many
light multiquark hadrons above 1~GeV? Also the existence of
$\pi_1$-mesons (among them the $\pi_1(1400)$ and $\pi_1(1600)$ are
well established~\cite{pdg}) is not compatible with the standard
quark model. Usually the $\pi_1$-mesons are interpreted as some
hybrid quark-gluon states. A question then appears why do not we
see other hybrid states with exotic quantum numbers, e.g. with
$J^{PC}=0^{--}$ or $2^{--}$, among light mesons? There are some
deep theoretical questions as well. The light hadrons represent
highly relativistic quantum systems in which $L$ is not a
conserved quantity.
% and the usual quantum-mechanical rule for the total spin, $J=L+S$, is
% questionable.
The quark angular momentum $L$ is nevertheless a standard
ingredient in constructing dynamical models for light hadrons.
Another question concerns the observable quantities like hadron
masses --- they must be renorminvariant in the field-theoretical
sense. Stated simply, they must represent some constants
independent of energy scale. The quark masses, for instance, are
not such constants since they have anomalous dimension in QCD. A
relation of calculated observables to the renormalization
invariance is obscure in almost all phenomenological models of
hadrons.

In the present work, we put forward a principally new approach to
the light hadrons in which the masses are relativistic and
renorminvariant by construction. The approach allows to classify
the light mesons without use of any angular momentum associated
with hadron constituents. The states $\pi_1$ emerge in a natural
way while the other exotic quantum numbers remain forbidden. The
constructed mass counting scheme permits to obtain hadron masses
from very simple relations with a typical accuracy comparable to
numerical calculations in complicated dynamical models.

%Below we first motivate our approach and apply it to the light
%mesons composed of the $u$ and $d$ quarks. Then we discuss a
%complementary and likely more realistic interpretation (at least
%below 1~GeV) of the model and incorporate the strange quark.
%Extensions to baryons and to heavy flavors will be considered
%elsewhere\footnote{We plan to address this issue in a follow-up
%paper.}.

\section{Motivation and idea of the approach}

Consider the Gell-Mann--Oakes--Renner relation for pion
mass~\cite{gor},
\begin{equation}
\label{1}
m_\pi^2=-\frac{\langle\bar{q}q\rangle}{f_\pi^2}(m_u+m_d)=\Lambda\cdot2m_q,
\end{equation}
where we set $m_u=m_d\doteq m_q$ and
$\Lambda\doteq-\frac{\langle\bar{q}q\rangle}{f_\pi^2}$. We will
use the standard values for the quark condensate and masses of
current quarks at the scale of the pion mass from the QCD sum
rules~\cite{svz} and Chiral Perturbation Theory
(ChPT)~\cite{gasser},
$\langle\bar{q}q\rangle=-(250\,\text{MeV})^3$, $m_u+m_d=11$~MeV.
Together with $f_\pi=92.4$~MeV (the value of the weak pion decay
constant in the normalization used in~\eqref{1}), the
relation~\eqref{1} yields $m_\pi=140$~MeV and $\Lambda=1830$~MeV.
The famous relation~\eqref{1} can be derived in various ways as a
consequence of the Spontaneous Chiral Symmetry Breaking (SCSB) in
strong interactions. The interpretation of pion as the
pseudogoldstone boson arising due to the SCSB in QCD became a
common lore in particle physics.

Let us look at the relation~\eqref{1} from a different angle.
According to Quantum Mechanics, when a fermion-antifermion pair
has the lowest energy it carries the quantum numbers of a
pseudoscalar particle. This is the standard description of ground
pseudoscalar particles in the hadron potential models where the
notion of SCSB is absent. They have problems, however, with small
pion mass and ensuing low-energy physics. We believe that the
quantum-mechanical intuition about the pion is right and the
problem with small pion mass appears because the relativistic and
renormalization invariance of underlying theory are not properly
reflected even in "relativized" potential models. If this
viewpoint is correct, there should exist an alternative way for
derivation of relation~\eqref{1}, a way which does not use the
assumption of SCSB. We do not have such a derivation in a rigorous
way but we can outline a possible direction one should pursue to
get it.

The mass of a hadron state $|h\rangle$ can be related to the trace
of energy momentum tensor in QCD via the following Ward identity
known in deep inelastic phenomenology~\cite{jaffe},
\begin{equation}
\label{2}
2m_h^2=\langle h|\Theta^\mu_\mu|h\rangle,
\end{equation}
where $\Theta^\mu_\mu$ is given by the scale anomaly,
\begin{equation}
\label{3}
\Theta^\mu_\mu=\frac{\beta}{2g_s}G_{\mu\nu}^2+\!\!\sum_{\psi=u,d,\dots}\!\!\!\! m_\psi\bar{\psi}\psi.
\end{equation}
Here $\beta$ denotes the QCD Beta-function. As follows from
derivation of the identity~\eqref{2}, the squared mass (not the
linear one!) in the l.h.s. appears due to the relativistic
invariance and the general factor 2 stems from the symmetry
between particles and antiparticles in relativistic Hamiltonians.
The r.h.s. of~\eqref{2} represents a renorminvariant quantity. One
can build two substantially different renormalization invariant
operators in QCD, the both are present in~\eqref{3}. Applying the
identity~\eqref{2} to $\pi$-meson, one should prove that
$\langle\pi|G_{\mu\nu}^2|\pi\rangle=0$ and
$\langle\pi|\bar{q}q|\pi\rangle=\frac{2}{f_\pi^2}\langle0|\bar{q}q|0\rangle$,
$q=u,d$. This leads to the relation~\eqref{1}.

We assume thus that the relation~\eqref{1} is a result of the Ward
identity~\eqref{2} applied to the lightest hadron. The final mass
formula has a form enabling the interpretation based on the SCSB.
The identity~\eqref{2} suggests in this case that masses of other
hadrons composed of $u$ and $d$ quarks can be given by extensions
of~\eqref{1} in which $\langle h|G_{\mu\nu}^2|h\rangle\neq0$.
Namely,
\begin{equation}
\label{4}
m_h^2=\Lambda(E_h+2m_q)=\Lambda E_h+m_\pi^2.
\end{equation}
The quantity $E_h\sim\langle h|G_{\mu\nu}^2|h\rangle\neq0$ depends
on a hadron under consideration, with the combination $\Lambda
E_h$ being renorminvariant. As hadron masses are inseparably
connected with properties of non-perturbative QCD vacuum, there
should be a universal way to relate $E_h$ to the gluon
condensate\footnote{We plan to address this issue in a follow-up
paper.}, $E_h\sim\langle 0|G_{\mu\nu}^2|0\rangle\neq0$, at least
in the chiral limit. In the ansatz~\eqref{4}, we take into account
that "switching off" the explicit contribution from gluons we must
come to the pion mass.

The simultaneous restrictions imposed by the relativistic and
renormalization invariance are likely so strong that we do not
have a large choice in constructing expressions for hadron
masses\footnote{The vast majority of hadron resonances have no
well defined gravitational (and inertial) mass as they do not
propagate in space: Their typical lifetime of the order of
$10^{-23}$--$10^{-24}$s does not allow to leave the reaction zone
of the order of 1~fm which is their size simultaneously. They show
up only as some structures in the physical observables at certain
energy intervals. The resonance mass is commonly associated with
the real part of an $S$-matrix pole on the second (unphysical)
sheet. How this definition is related with the gravitational mass
is by far not obvious in a theory with confinement. The Ward
identity~\eqref{2} can be considered as a definition of
gravitational hadron mass that extends naturally to the masses of
unstable resonances (assuming the existence of a limit where their
gravitational masses are well defined).}.
%The ansatz~\eqref{4} is perhaps the only one for the "standard"
%light non-strange mesons.
Below we develop a simple model based on the ansatz~\eqref{4} that
describes the whole spectrum of light mesons.
%these hadrons.

The first excitation of pion is the $\rho$-meson. In hadron
reactions with pions, this excitation can be interpreted as a
quark spin flip caused by absorption of a long-distance gluon.
Within the QCD sum rules, such non-perturbative processes are
parametrized via a contribution of the gluon condensate $\langle
0|G_{\mu\nu}^2|0\rangle\doteq \langle
G_{\mu\nu}^2\rangle$~\cite{svz}. The given spin flip "costs"
energy $E_\rho\approx310$~MeV in our model. The ansatz~\eqref{4}
yields then\footnote{The Particle Data reports
$m_\rho=775.11\pm0.34$~MeV for the charged
$\rho$-meson~\cite{pdg}. This averaged mass is seen in leptonic
processes~\cite{pdg}. In the hadronic processes, however, the
averaged mass of charged $\rho$ is
$m_\rho=766.5\pm1.1$~MeV~\cite{pdg}. The difference can be easily
explained: One needs an extra energy to create two current quarks
in the leptonic processes and usually does not need this in the
hadronic reactions. The latter situation is our case.}
$m_\rho\approx766.5$~MeV. The crucial point here is that the
non-relativistic logic (or rather the "non-renorminvariant" one)
$m_\rho=E_\rho+m_\pi$ does not work\footnote{A remote analogy: The
non-relativistic addition law for velocities, $v=v_1+v_2$, does
not work in the relativistic case and must be replaced by a
non-linear law.} because $E_\rho$ is not renorminvariant: The
value of $310$~MeV refers to the scale where $\Lambda=1830$~MeV.
The relativistic and renorminvariant relation between $m_\rho$ and
$E_\rho$ is given by our ansatz~\eqref{4}. Under some assumptions,
$E_\rho$ can be expressed via the values of vacuum quark and gluon
condensates\footnote{See the footnote~1.},
$E_\rho=\frac{\beta_0}{24}\frac{(\alpha_s/\pi)\langle
G_{\mu\nu}^2\rangle}{|\langle\bar{q}q\rangle|}$, where
$\beta_0=11-\frac23n_f$ and we take $\frac{\alpha_s}{\pi}\langle
G_{\mu\nu}^2\rangle=0.012(3)$~GeV$^4$~\cite{svz}. For $n_f=2$, the
value $E_\rho\approx310$~MeV is reproduced for the center value of
the gluon condensate. In the present phenomenological model,
however, we will treat $E_\rho$ as an input parameter, this is
enough for our purposes. It is interesting to observe that the
value of $E_\rho$ resembles a typical value of constituent quark
mass\footnote{Strictly speaking, the value of constituent quark
mass is very model dependent --- as far as we know, it ranges from
220 to 450~MeV in various models. The value
$M_c(p^2)\simeq310$~MeV at small momentum $p$, however, proves to
be seen in unquenched lattice simulations in the chiral limit
(see, e.g., Ref.~\cite{bowman}). We may indicate a "pocket" way
for reproducing this value. When one incorporates the vector and
axial mesons into the low-energy models describing the SCSB in QCD
and related physics, e.g. into the Nambu--Jona-Lasinio
model~\cite{klev} or the linear sigma-model, a model-independent
relation emerges: $m_{a_1}^2=m_\rho^2+6M_c^2$. On the other hand,
the idea of SCSB was also exploited in the derivation of famous
Weinberg relation, $m_{a_1}^2=2m_\rho^2$~\cite{wein}. Combining
both relations, one has $M_c=m_\rho/\sqrt{6}$ that gives the
aforementioned value.}, $M_c\approx m_p/3$, where $m_p$ is the
proton mass.

\section{Meson excitations above 1~GeV}

We assume that the masses of higher spin and radial excitations
are relativistic and renorminvariant quantities and can be given
by the ansatz~\eqref{4}. The description with correct quantum
numbers can be realized within the following scheme. We conjecture
that gluodynamics leads to formation of isoscalar quasiparticles
$A_0$ and $A_1$ which carry the pseudoscalar, $J^{PC}=0^{-+}$, and
vector, $J^{PC}=1^{--}$, quantum numbers, respectively. They
represent gluonic counterparts of para- and ortho-positronium ---
the massive states that can be formed in photonic processes and
"dematerializing" into pure photons. The quasiparticles $A_0$ and
$A_1$, generally speaking, are not necessarily colorless. The
underlying nature of these objects is not important for our
further purposes. They reside likely under the valent $q\bar{q}$
shell and we will call them "underquarkonia" in what follows. The
creation of underquarkonia requires energy $E_0$ for $A_0$ and
$E_1$ for $A_1$. By assumption, these energies are proportional to
the gluon condensate $\langle G_{\mu\nu}^2\rangle$, so that the
resulting products $\Lambda E_0$ and $\Lambda E_1$ are
renorminvariant. The energies $E_0$ and $E_1$ are "building
blocks" which constitute the total excitation energy $E_h$
in~\eqref{4}. Within this scheme, the meson mass depends on the
total number of underquarkonia inside the given meson. It is
convenient to divide the non-strange resonances above 1~GeV into
excited pions and excited $\rho$($\omega$) mesons. The excited
$\rho$ mesons containing $n$ underquarkonia $A_0$ and $l$
underquarkonia $A_1$ will be labeled
% $^J\!{\mathlarger{\mathlarger{\boldsymbol{\rho}}}}_{{A_0^n A_1^l}}$
$^J\!{\mathlarger{\mathlarger{\rho}}}_{{\!A_0^n A_1^l}}$, where $J$ means
the total spin. The mass of these excitations is given by
\begin{equation}
\label{5}
m_{{\mathlarger{\mathlarger{\rho}}}_{{\!A_0^n A_1^l}}}^2=\Lambda(E_\rho+nE_0+lE_1)+m_\pi^2.
\end{equation}
By assumption, the nucleus formed by underquarkonia is in the
ground $S$-wave state. This leads to the maximal spin $J=l+1$, the
$P$ and $C$ parities obey to the standard multiplicative law:
$(P,C)=((-1)^{n+l+1},(-1)^{l+1})$. The excited pions follow the
same principle: The states
%$^J{\mathlarger{\mathlarger{\boldsymbol{\pi}}}}_{{\! A_0^n A_1^l}}$
$^J{\mathlarger{\mathlarger{\pi}}}_{{\!\! A_0^n A_1^l}}$ have mass
\begin{equation}
\label{5b}
m_{{\mathlarger{\mathlarger{\pi}}}_{{\!\!A_0^n A_1^l}}}^2=\Lambda(nE_0+lE_1)+m_\pi^2,
\end{equation}
maximal spin $J=l$, and parities $(P,C)=((-1)^{n+l+1},(-1)^l)$.

Since the pair $A_0A_0$ does not change neither spin nor parities,
a meson containing even number $2k$ of $A_0$ and no one $A_1$ will
be referred to as the $k$-th radial excitation of the ground
state, by definition. The radial excitations appear thus with a
"period" $2E_0\Lambda$. The value of $E_0$ can be fixed from
masses of pion radial excitations $\pi(1300)$ and $\pi(1800)$
which are $^0{\mathlarger{\mathlarger{\pi}}}_{{\!\! A_0^2}}$ and
$^0{\mathlarger{\mathlarger{\pi}}}_{{\!\! A_0^4}}$ states in our
notations. This fixes\footnote{We used $\pi(1800)$ as the
uncertainty in its mass is much smaller than in the case of
$\pi(1300)$~\cite{pdg}.} $E_0\approx440-450$~MeV. $E_1$ is fixed
in the next Section.

It is easy to see that the excitations of the kind
$^J\!{\mathlarger{\mathlarger{\rho}}}_{{\!A_1^{2k}}}$,
$k=0,1,2,\dots$, will give rise to a degenerate family of
resonances with $J^{PC}=(1,3,\dots,2k+1)^{--}$ (the even spins do
not appear as explained in the next Section). The states on the
main $\rho$-meson Regge trajectory follow thus with a "period"
$2E_1\Lambda$. These excitations, in addition, generate daughter
trajectories. For instance, the excited $\rho$ mesons emerge with
$J^{PC}=1^{--}$ which follow with the same "period" $2E_1\Lambda$.
They are different from the "radial" $\rho$ mesons following with
another period $2E_0\Lambda$.

Let us summarize the constructed picture for the excited light
non-strange mesons above 1~GeV. A meson looks like an atom.
Strictly speaking, it looks like a unusual positronium atom
containing usual atomic nucleus. The role of electron and positron
is played  by valent quark and antiquark in the spin-singlet
("para") or spin-triplet ("ortho") states. The role of neutron and
proton is played  by the para- and ortho-underquarkonia. The mass
of the usual atom, in the first approximation, is just sum of
masses of all nucleons and electrons. In the case of hadronic
atom, this mass counting rule contradicts both the relativistic
and renormalization invariance. It must be replaced by the
rule~\eqref{5} for the ortho- and by~\eqref{5b} for the
para-states. The spins of the ortho-underquarkonia must be added
as spins of massless particles (see the next Section). Following
these prescriptions, one can construct a "periodic table of the
hadronic elements" describing the masses and quantum numbers of
light non-strange mesons. In real situations, however, this
idealistic picture is contaminated by contributions of the strange
quark. A careful systematization requires a thorough
phenomenological study of available data which we leave for a
future work.

If underquarkonia are interpreted as real quark-antiquark pairs
then the hadron excitations containing $A_0$ and $A_1$ represent
multiquark states.

\section{Some examples of meson excitations}

Below we dwell on some meson excitations to demonstrate our
scheme.

The state $^0{\mathlarger{\mathlarger{\pi}}}_{{\!\! A_0}}$ has the
quantum numbers of scalar particle, $J^{PC}=0^{++}$. Its mass is
$m=\sqrt{\Lambda E_0+m_\pi^2}\approx910$~MeV. It could be
$a_0(980)$ in which contributions from the strange quarks were not
taken into account.

$^1\!{\mathlarger{\mathlarger{\rho}}}_{{\!A_0}}$ and
$^1{\mathlarger{\mathlarger{\omega}}}_{{\!A_0}}$ have quantum
numbers $J^{PC}=1^{+-}$ and mass about 1180~MeV. The natural
candidates are $h_1(1170)$ and $b_1(1230)$~\cite{pdg}.

$^J\!{\mathlarger{\mathlarger{\rho}}}_{{\!A_1}}$ and
$^J{\mathlarger{\mathlarger{\omega}}}_{{\!A_1}}$ have
$J^{PC}=(0,1,2)^{++}$. They should be the series of states
$a_1(1260)$, $f_1(1285)$, $f_2(1270)$, $a_2(1320)$, and likely
$f_0(1370)$~\cite{pdg}. Fitting to the masses of well measured
resonances $f_1(1285)$ and $f_2(1270)$, we obtain the estimate
$E_1\approx570$~MeV which will be used in what follows.

$^1\!{\mathlarger{\mathlarger{\rho}}}_{{\!A_0^2}}$ and
$^1{\mathlarger{\mathlarger{\omega}}}_{{\!A_0^2}}$ of mass near
1490~MeV are first radial excitations of $\rho$ and $\omega$
mesons --- the resonances $\rho(1450)$ and
$\omega(1420)$~\cite{pdg}.

$^J\!{\mathlarger{\mathlarger{\rho}}}_{{\!A_0A_1}}$ and
$^J{\mathlarger{\mathlarger{\omega}}}_{{\!A_0A_1}}$ have quantum
numbers $J^{PC}=(0,1,2)^{-+}$, i.e. 3 possible spins and pion
parities. These states describe resonances $\pi$, $\pi_1$ and
$\pi_2$ above 1~GeV. In contrast to the quark model, the
$\pi_1$-meson is not exotic in our scheme! The relation~\eqref{5b}
predicts mass about 1570~MeV. The possible candidates are
$\pi_1(1600)$ and $\pi_2(1670)$~\cite{pdg}. In the real hadron
reactions, however, it seems that the pseudoscalar underquarkonium
$A_0$ is sometimes replaced by the lighter $\pi^0$ meson. This may
happen when photons come into play and participate in formation of
underquarkonia along with gluons. In the case under consideration,
a lighter state
$^J\!{\mathlarger{\mathlarger{\rho}}}_{{\!\pi^0\!A_1}}$ with mass
near 1370~MeV can be formed instead of
$^J\!{\mathlarger{\mathlarger{\rho}}}_{{\!A_0A_1}}$. The resonance
$^1\!{\mathlarger{\mathlarger{\rho}}}_{{\!\pi^0\!A_1}}$ in
particular is a candidate for $\pi_1(1400)$ (experimentally,
$m_{\pi_1(1400)}=1354\pm25$~MeV~\cite{pdg}). The state
$^0\!{\mathlarger{\mathlarger{\rho}}}_{{\!\pi^0\!A_1}}$ is another
candidate for the radially excited pion. This may mean that
$\pi(1300)$ has two possible structures ---
$^0{\mathlarger{\mathlarger{\pi}}}_{{\!\! A_0^2}}$ and
$^0\!{\mathlarger{\mathlarger{\rho}}}_{{\!\pi^0\!A_1}}$ --- that
could be a reason for a large uncertainty in the mass of
$\pi(1300)$, $m_{\pi(1300)}=1300\pm100$~MeV~\cite{pdg}. In
general, multiple possibilities for formation of a hadron
resonance become rather common for highly excited states. This
could be among reasons of the observed tendency of growing total
width as a function of mass.

The states $^J\!{\mathlarger{\mathlarger{\rho}}}_{{\!A_1^2}}$ and
$^J{\mathlarger{\mathlarger{\omega}}}_{{\!A_1^2}}$ have
$J^{PC}=J^{--}$. A new feature appears in these states: We must
add spins of two ortho-underquarkonia $A_1$. The standard spin
counting rule would give $J=0,1,2,3$ for the total hadron spin.
But our spin addition law will be different. To motivate this new
rule consider first two photons. The photon is massless, hence,
its spin has two projections only --- two possible helicities. The
total helicity of two photons can be 0 or 2. In other words, the
addition law for spins of spin-1 massless particles is the same as
for massive spin-$\frac12$ particles, just the final result must
be multiplied by 2. Consider now an ortho-positronium which decays
into 3 photons (decay into 2 photons is forbidden). If 2
ortho-positronia are simultaneously created by 6 photons, their
total spin can be 0 or 2 because the total helicity of original
photons is even\footnote{The total spin of ortho-positronia can be
"rotated" to the value of 1 by some subsequent manipulations.}. We
come thus to the conclusion that although ortho-positronia are
massive spin-1 particles, their spins must be added (immediately
after formation) as if they were massless. The same rule we apply
to gluons and underquarkonia. The total spin of the pair $A_1A_1$
can be 0 or 2, hence, the total spin of
$^J\!{\mathlarger{\mathlarger{\rho}}}_{{\!A_1^2}}$ can be $J=1,3$.
A natural candidate is the couple of $J^{PC}=(1,3)^{--}$ states
$\rho(1700)$ and $\rho_3(1690)$ (and $\omega(1650)$ with
$\omega_3(1670)$ for $\omega$)~\cite{pdg}. Our results do not
contradict the quark model predictions. First, the exotic states
with quantum numbers $J^{PC}=(0,2)^{--}$ are absent. Second,
$\rho(1700)$ is not the second radial excitation\footnote{In the
quark potential models, $\rho(1700)$ (and $\omega(1650)$) is a
$D$-wave state, while the radial excitations of $\rho(770)$ should
be $S$-wave ones.} of $\rho(770)$.

The recipe for constructing higher radial and spin excitations is
straightforward. An interesting question emerge: Can the
underquarkonia $A_0$ and $A_1$ materialize as real hadrons with
pseudoscalar and vector quantum numbers? If this happens, the
masses of pure underquarkonia should be estimated as
$M_{A_i}=\sqrt{\Lambda E_i}$, $i=0,1$, according to our ansatz.
The mass of pseudoscalar "materialized" underquarkonium,
$M_{A_0}\approx910$~MeV, is suspiciously close to the mass of
$\eta'$-meson, $m_{\eta'}=958$~MeV~\cite{pdg}. The vector
underquarkonium would materialize with the mass
$M_{A_1}\approx1020$~MeV which coincides with the $\varphi$-meson
mass. These observations suggest that the idea of underquarkonia
might be able to give an alternative non-instanton solution for
the old $U(1)_A$ problem, i.e. to explain why $\eta'$ is much
heavier than it is prescribed by the quark model, and to clarify
why the strange component in $\varphi$ is almost unmixed with the
lighter quarks.

\section{Regge trajectories}

The relations~\eqref{5} and~\eqref{5b} lead to linear Regge,
equidistant daughter Regge and radial trajectories. Below we give
examples for some of them.

The states $^0{\mathlarger{\mathlarger{\pi}}}_{{\!\! A_0^{2n}}}$,
$n=0,1,2,\dots$, form linear trajectory for the radial excitations
of pion,
\begin{equation}
\label{6}
m_\pi^2(n)=2\Lambda E_0n+m_\pi^2.
\end{equation}
The radial excitations of $\rho(770)$ --- the resonances
$^1\!{\mathlarger{\mathlarger{\rho}}}_{{\!A_0^{2n}}}$ --- lie on
the first radial $\rho$-trajectory,
\begin{equation}
\label{7}
m_\rho^2(n)_I=2\Lambda E_0\left(n+\frac{E_\rho}{2E_0}\right)+m_\pi^2.
\end{equation}
The second radial $\rho$-trajectory is composed of the states
$^1\!{\mathlarger{\mathlarger{\rho}}}_{{\!A_1^2A_0^{2n}}}$,
\begin{equation}
\label{8}
m_\rho^2(n)_{I\!I}=2\Lambda E_0\left(n+\frac{E_1}{E_0}+\frac{E_\rho}{2E_0}\right)+m_\pi^2.
\end{equation}
It is evident that the states
$^1\!{\mathlarger{\mathlarger{\rho}}}_{{\!A_1^{2(\!k-1\!)}A_0^{2n}}}$
formally give rise to the $k$-th radial $\rho$-trajectory. The
resonances having structure
$^1\!{\mathlarger{\mathlarger{\rho}}}_{{\! A_1A_0^{2n}}}$ form the
first radial $a_1$-trajectory,
\begin{equation}
\label{9}
m_{a_1}^2(n)_I=2\Lambda E_0\left(n+\frac{E_1}{2E_0}+\frac{E_\rho}{2E_0}\right)+m_\pi^2.
\end{equation}
The further axial radial trajectories can be easily written. It
should be remarked that the standard quark model allows of only
two radial $\rho$-trajectories ($S$- and $D$-wave ones) and one
$a_1$-trajectory (a $P$-wave one). Our scheme is richer.

The spin $\rho$-trajectory is composed from the states
$^J\!{\mathlarger{\mathlarger{\rho}}}_{{\!A_1^{J-1}}}$ with
$J=1,3,5,\dots$. The corresponding masses are
\begin{equation}
\label{10}
m_{\rho_J}^2=\Lambda E_1\left(J-1+\frac{E_\rho}{E_1}\right)+m_\pi^2.
\end{equation}
For the even spins, $J=2,4,6,\dots$, the trajectory~\eqref{10}
describes $a_J$ ($f_J$) mesons. The Regge trajectory~\eqref{10}
describes thus states with alternating parities and this agrees
with the phenomenology.

The slopes of radial trajectories~\eqref{6}--\eqref{9} are
universal. This property was observed in the phenomenology of
light mesons~\cite{phen}. The slope $2\Lambda E_1$ of spin Regge
trajectory~\eqref{10} is different from the radial one as it is
determined by the energy $E_1$ of different underquarkonium. If we
consider all spins in~\eqref{10} then the slope is $\Lambda
E_1\approx1.83\cdot0.57\approx1.04$~GeV$^2$. This value is in
accord with the phenomenology~\cite{phen}. The phenomenological
value of Regge slope can be taken, in principle, from the
experimental data to fix $E_1$.

The principal $\rho$-meson Regge trajectory~\eqref{10} is
accompanied by the daughter trajectories following with the step
$2\Lambda E_1$. The spin-1 $\rho$-mesons of the kind
$^1\!{\mathlarger{\mathlarger{\rho}}}_{{\!A_1^{J-1}}}$ are the
lowest states on the daughters. For example, the lowest state
in~\eqref{8} is the lowest state on the first daughter. The
spectrum of $^1\!{\mathlarger{\mathlarger{\rho}}}_{{\!A_1^{J-1}}}$
excitations reads
\begin{equation}
\label{10b}
m_{^1\!{\mathlarger{\mathlarger{\rho}}}_{{\!A_1^{J-1}}}}^2=
2\Lambda E_1\left(k+\frac{E_\rho}{2E_1}\right)+m_\pi^2,
\end{equation}
where $k=0,1,2,\dots$ enumerates the daughters. Similar relations
can be written for the axial and other mesons. An obvious
consequence of the emerging spectrum is the degeneracy of spin and
daughter radial excitations of the type $m^2(J,k)\sim J+k$, which
is typical for the Veneziano dual amplitudes~\cite{avw} and the
Nambu--Goto open strings.

The values of $E_\rho$, $E_0$ and $E_1$ are of similar magnitude.
This allows to consider a reasonable limit where they are equal.
The notions of radial and daughter radial trajectories coincide in
the limit $E_0=E_1$. In the limit $E_\rho=E_1$, the radial vector
and axial trajectories are related by
\begin{equation}
\label{11}
m_{a_1}^2(n)=m_\rho^2(n)+m_\rho^2-m_\pi^2.
\end{equation}
This relation holds both for radial and for daughter radial
trajectories. In the chiral limit, $m_\pi=0$, the
relation~\eqref{11} for the ground states, $n=0$, reduces to the
old Weinberg relation, $m_{a_1}^2=2m_\rho^2$~\cite{wein}. In the
most symmetric limit, $E_\rho=E_0=E_1$, the vector and axial
radial spectrum reduce to a simple form in the chiral limit,
\begin{equation}
\label{12}
m_\rho^2(n)=2m_\rho^2\left(n+\frac12\right),
\end{equation}
\begin{equation}
\label{13}
m_{a_1}^2(n)=2m_\rho^2\left(n+1\right),
\end{equation}
The relations~\eqref{12} and~\eqref{13} first appeared in the
variants of Veneziano amplitude which incorporated the Adler
self-consistency condition~\cite{avw}. This condition (the
amplitude of $\pi\pi$ scattering is zero at zero momentum) removes
degeneracy between the $\rho$ and $a_1$ spectra. Within the QCD
sum rules, the relations~\eqref{12} and~\eqref{13} may be
interpreted as a large-$N_c$ generalization of the Weinberg
relation~\cite{afonin:plb}.

We see thus that in certain limits the Regge
phenomenology of our approach reproduces various known relations.

Let us clarify further how the excited resonances with identical
quantum numbers can have different origin in the proposed scheme.
They may represent the radial states, states on daughter Regge
trajectories and various "mixed" ones. For instance, the second
$\rho$-meson excitation with the same parities (i.e. containing 2
pairs of underquarkonia) appears in three forms: the second radial
excitation $^1\!{\mathlarger{\mathlarger{\rho}}}_{{\!A_0^4}}$, the
vector state on the second daughter trajectory
$^1\!{\mathlarger{\mathlarger{\rho}}}_{{\!A_1^4}}$, and the mixed
one $^1\!{\mathlarger{\mathlarger{\rho}}}_{{\!A_0^2A_1^2}}$. They
are degenerate only in the limit $E_0=E_1$. It is likely difficult
to detect such a splitting experimentally because of overlapping
widths. In reality, one would observe rather a "broad resonance
region".

We note also that placing the observed "radial" states on a
certain trajectory should be made with care --- an incorrect
interpretation of states leads to a false (or more precisely,
introduced by hands) non-linearity of the trajectory. Take again
the $\rho$ meson as a typical example. The first 3 radial
excitations of $\rho$ are the states
$^1\!{\mathlarger{\mathlarger{\rho}}}_{{\!A_0^2}}$,
$^1\!{\mathlarger{\mathlarger{\rho}}}_{{\!A_0^4}}$, and
$^1\!{\mathlarger{\mathlarger{\rho}}}_{{\!A_0^6}}$. They are
accompanied by the following states with the quantum numbers of
$\rho$: $^1\!{\mathlarger{\mathlarger{\rho}}}_{{\!A_1^2}}$,
$^1\!{\mathlarger{\mathlarger{\rho}}}_{{\!A_1^2A_0^2}}$, and
$^1\!{\mathlarger{\mathlarger{\rho}}}_{{\!A_1^4}}$. Since $\frac32
E_0\!>\!E_1\!>\!E_0$ in our fits, the sequence of first 7
$\rho$-mesons is: $\mathlarger{\mathlarger{\rho}}$,
$^1\!{\mathlarger{\mathlarger{\rho}}}_{{\!A_0^2}}$,
$^1\!{\mathlarger{\mathlarger{\rho}}}_{{\!A_1^2}}$,
$^1\!{\mathlarger{\mathlarger{\rho}}}_{{\!A_0^4}}$,
$^1\!{\mathlarger{\mathlarger{\rho}}}_{{\!A_1^2A_0^2}}$,
$^1\!{\mathlarger{\mathlarger{\rho}}}_{{\!A_1^4}}$,
$^1\!{\mathlarger{\mathlarger{\rho}}}_{{\!A_0^6}}$. They likely
correspond to the vector resonances $\rho(770)$, $\rho(1450)$,
$\rho(1700)$, $\rho(1900)$, $\rho(2000)^*$, $\rho(2150)$,
$\rho(2270)^*$. Here the states marked by asterisk are taken from
"Further States" in the Particle Data~\cite{pdg}.

\section{$\sigma$-meson as a collisional excitation}

The valent electron in atom can be excited in two ways: (a) via
absorption of photon (with ensuing spin flip due to momentum
conservation); (b) via a direct collision with some particle.
Looking at an excited atom at some moment in time $t_0$ we cannot
say how it was excited without any information got prior to the
moment $t_0$. There exists the third way for changing energy of
atomic electron: (c) via a change of nucleus charge caused, e.g.,
by absorption of $\alpha$-particle.

Let us apply these electromagnetic analogies to the pion
excitations. Within our approach, the case (a) corresponds to the
$\rho$-meson. The case (c), in some sense, corresponds to the
formation of underquarkonia. But what about the option (b)? In
practical situations of collisions of hadronic beams, the
"collisional" excitations should play an important or even very
important role.

Consider a low-energy $\pi\pi$ scattering. We may have a situation
when a pion collides as a whole with one of quarks of another
pion. This may happen when the first pion is faster, hence, has
smaller de Broglie wavelength, i.e. when one pion wave packet
penetrates another one\footnote{A picture of this process depends
on a reference frame.}. The collision lasts a very short time
$\Delta t$ which determines the lifetime of the formed coherent
state. During the time $\Delta t$, the second quark
(quark-spectator) "feels" the first one as a particle with
unchanged color charge and spin (because pion is colorless and
spinless) but with different proper mass: $m_q\rightarrow
m_q+m_\pi$. According to our basic principle~\eqref{4}, this
coherent state, let us denote it $\sigma$ beforehand, obeys to the
mass relation
\begin{equation}
\label{14}
m_\sigma^2=\Lambda m_\pi + m_\pi^2,
\end{equation}
that yields $m_\sigma\approx525$~MeV. This state must have the
quantum numbers of scalar meson, decay into two pions, and should
be an extremely short-living particle. The formation of such a
coherent state is favored by Coulomb attraction between $\pi^+$
and $\pi^-$ which leads to
%(and not in $\pi^{\pm}\pi^0$ and $\pi^0\pi^0$ cases).
the dominance of isosinglet channel. All these observations
strongly suggest that we must interpret this resonance as the
$f_0(500)$-meson~\cite{pdg}, widely known as $\sigma$-meson.

We may propose a quantum-mechanical interpretation for the
relation~\eqref{14}. Consider a stationary state of some quantum
system with energy $E$ which is described by the Schr\"{o}dinger
equation, $H|\psi\rangle=E|\psi\rangle$. Consider now a sudden
perturbation of this system, e.g. by a very fast particle. A
sudden perturbation in Quantum Mechanics is a perturbation lasting
so short time $\Delta t$ that the wave function $\psi$ does not
change during $\Delta t$ (such a change always requires a finite
time). The perturbed system is described by the equation
$H'|\psi\rangle=E'|\psi\rangle$. But $\psi$ is not an
eigenfunction of the new Hamiltonian $H'$ and the perturbed system
becomes unstable. An important point for us is that the form of
$\psi$ determines the functional dependence of energy $E$ from the
parameters $a_i$ of the Hamiltonian $H$, $E=E(a_i)$. Since $\psi$
is unchanged during the time $\Delta t$, this dependence remains
the same after the perturbation. It means that if $H'$ is
determined by a set of perturbed parameters $a'_i$ then
$E'=E(a'_i)$. Returning to the $\pi\pi$ scattering, one of pion
can be considered as a stationary system in which $m_q$ plays the
role of a parameter determining its energy $m_\pi$. A collision
causes a sudden perturbation that changes the parameter,
$m_q\rightarrow m_q+m_\pi$, for a short time $\Delta t$. This
change must be substituted to~\eqref{1} (for one of quarks) to
obtain the energy~\eqref{14} of unstable state $\sigma$.

Within our approach, the main difference of a collisional
resonance from the resonances considered above consists in the
absence of explicit renorminvariance of its mass: The product
$\Lambda m_\pi$ in~\eqref{14} is not renorminvariant and $\Lambda$
should be taken at the scale of $m_\pi$. If we replace pion
in~\eqref{14} by some other particle, we should take $\Lambda$ at
the scale of that particle.

The recent lattice simulation of Ref.~\cite{briceno} reported an
effect of evolving $\sigma$-meson into a stable bound state lying
below the $\pi\pi$ threshold as $m_\pi$ is increased. This
observation follows directly from the mass relation~\eqref{14}:
Imposing $m_\sigma\geqslant2m_\pi$ we obtain
$m_\pi\leqslant\Lambda/3$. This restriction is nontrivial as
$\Lambda$ depends on the mass scale. If we normalize $\Lambda$ to
the numerical results of Ref.~\cite{briceno}, $m_\sigma=758$~MeV
when $m_\pi=391$~MeV, we get $\Lambda=1078$~MeV that gives the
estimate $m_\pi\leqslant359$~MeV. This restriction agrees with the
lattice results of Ref.~\cite{briceno}: $\sigma$ represents a
bound state at $m_\pi=391$~MeV and a broad resonance at
$m_\pi=236$~MeV.

\section{$h_1$ and $b_1$ as collisional excitations}

In the potential quark models, the lightest scalar and
axial-vector mesons represent $P$-wave states and, as a result,
their masses are rather close since the difference stemming from
the spin-orbital interaction is relatively small. Within the
picture of "hadronic atom", we have identified the lightest axial
resonances $h_1$ and $b_1$ as
$^1\!{\mathlarger{\mathlarger{\rho}}}_{{\!A_0}}$ or
$^1{\mathlarger{\mathlarger{\omega}}}_{{\!A_0}}$ states. But we
can give also a collisional interpretation for theses states.
Imagine that a pion collided with a $\rho$-meson and this $\rho$,
having a smaller de Broglie wavelength, excited one of quarks
inside the pion in the collisional way. Let us denote the formed
coherent state $\pi_{\rho}$ (means "$\rho$ inside $\pi$"). The
$\sigma$-meson represents the $\pi_{\pi}$ collisional resonance in
this notation. According to our prescriptions, the mass of a
collisional state of the kind $\pi_{h}$ is given by the following
extension of relation~\eqref{14},
\begin{equation}
\label{15}
m_{\pi_{h}}^2=\Lambda m_h + m_\pi^2,
\end{equation}
where $\Lambda$ should be taken at the scale $m_h$. For making
estimates in the first approximation we will consider $\Lambda$ as
a universal constant and set as before $\Lambda=1830$~MeV. Taking
in~\eqref{15} $h=\rho$, we obtain $m_{h_1}\approx1190$~MeV that is
very close to our previous result. As in the $\sigma$-case, the
formation of the coherent state $\pi_{\rho}$ should be favored by
the Coulomb attraction, i.e. the favored channel is
$\pi_{\rho^-}^+$ or $\pi_{\rho^+}^-$ that entails zero isospin. A
natural consequence of $\pi_{\rho}$ structure of $h_1$ is the
absolute dominance of the decay mode
$h_1\rightarrow\rho\pi$~\cite{pdg}. The isotriplet partner of
$h_1(1170)$ --- the $b_1(1230)$ meson --- represents the
collisional resonance $\pi_{\omega}$ that determines its isospin~1
(it inherits the pion isospin) and dominant decay
$b_1\rightarrow\omega\pi$~\cite{pdg}. $b_1$ is expected to be
heavier than $h_1$ because $m_\omega>m_\rho$ and narrower than
$h_1$ because $\Gamma_\omega<\Gamma_\rho$. These expectations
agree with the experimental data~\cite{pdg}, at least
qualitatively.

It should be noted that the state
$^1{\mathlarger{\mathlarger{\pi}}}_{\! A_1}$ has the quantum
numbers of $b_1$ as well. The mass of this state is, however,
about\footnote{The state $^1{\mathlarger{\mathlarger{\eta}}}_{\!
A_1}$ nevertheless predicts the correct mass of 1170~MeV for the
$h_1(1170)$ meson (see the next Section).} 1030~MeV. We should
hence explain the suppression of this state within the atomic
picture --- a problem that does not arise in the collisional one.
The suppression seems to occur dynamically: As the energy $E_1$ is
accumulating inside a pion, the system prefers first to turn into
the $\rho$ meson since $E_\rho<E_1$ and the rest of energy
$E_1-E_\rho$ is not enough for further excitation of $\rho$.

We see thus that the axial $h_1$ and $b_1$ mesons can be
interpreted both as "hadronic atoms" and as collisional
resonances. This duality takes place due to an interesting
fine-tuning of parameters, $m_\rho\approx E_\rho+E_0$, that
provides an approximate coincidence of predicted masses in both
interpretations. Some (may be even all?) higher excitations allow
the dual interpretation as well. The collisional interpretation is
convenient for explaining dominant decay modes and isospin while
the atomic one suits better for deriving various mass relations
between different states, the Regge trajectories are a good
example.

\section{Towards inclusion of strange quark}

The ChPT predicts that all masses in the octet of pseudogoldstone
bosons follow from~\eqref{1} if one replaces $m_q$ by $m_s$ as
prescribed by the quark model~\cite{gasser}. An important point
here is that $\Lambda$ remains universal. We will use this
approximation to extend our mass counting scheme to light mesons
containing the strange quarks.

The quark content $q\bar{s}$ and
$(u\bar{u}+d\bar{d}-2s\bar{s})/\sqrt{6}$ of pseudoscalar $K$ and
$\eta$ mesons leads to the ChPT prediction for their masses:
$m_K^2=\Lambda(m_q+m_s)$ and $m_\eta^2=\Lambda(2m_q+4m_s)/3$. The
vector $K^*$ and $\varphi$ mesons have the quark content
$q\bar{s}$ and $s\bar{s}$, respectively. According to our
ansatz~\eqref{4}, the corresponding masses are
\begin{equation}
\label{16}
m_{K^*}^2=\Lambda(E_\rho+m_q+m_s) =  \Lambda E_\rho + m_K^2,
\end{equation}
\begin{equation}
\label{17}
m_\varphi^2=\Lambda(E_\rho+2m_s).
\end{equation}
The value of $m_s$ at the scale 1~GeV is about
$m_s\approx130$~MeV~\cite{pdg}. Accepting this value we obtain
$m_{K^*}\approx900$~MeV and $m_\varphi\approx1020$~MeV in a very
good agreement with their experimental values~\cite{pdg}.
Combining Eqs.~\eqref{16} and~\eqref{17} with~\eqref{5} for the
ground $\rho$-meson, we arrive at the Gell-Mann--Okubo mass
relation,
\begin{equation}
\label{18}
m_\rho^2+m_\varphi^2=2m_{K^*}^2.
\end{equation}
In the standard quark model, the relation~\eqref{18} holds for the
linear masses. The relativistic nature of our approach leads to
masses squared in~\eqref{18} automatically.

The higher spin and radial excitations can be built following the
identical scheme. The same applies to the collisional excitations.
For instance, the coherent state $K_{K^*}$ has mass $m_{K_{K^*}}=
\sqrt{\Lambda(m_q+m_s+m_{K^*})}\approx1380$~MeV, quantum numbers
of $h_1$, and decays into $K\bar{K^*}$ or $\bar{K}K^*$. This state
is observed as the resonance $h_1(1380)$~\cite{pdg}.

\section{The scalar sector below and near 1~GeV}

We have already interpreted $\sigma$-meson as the collisional
$\pi_\pi$ state and $h_1$($b_1$) as the $\pi_\rho$($\pi_\omega$)
one. Adding now the $K$ and $\eta$ mesons we can construct other
scalar collisional states which can be formed, e.g., in the $K\pi$
scattering.

Consider the state $\pi_{\mathsmaller{\!K}}$. Setting $h=K$
in~\eqref{15} we obtain its mass
$m_{{\mathlarger\pi}_{\mathsmaller{\!K}}}\approx970$~MeV. As in
the case of $\sigma$ and $h_1$, the expected isospin of
$\pi_{\mathsmaller{\!K}}$ is zero. Its natural decay mode would be
$\pi_{\mathsmaller{\!K}}\rightarrow K\pi$ but such a decay is
forbidden by the isospin conservation if $\pi_{\mathsmaller{\!K}}$
represents a genuine resonance. $\pi_{\mathsmaller{\!K}}$ should
be then relatively narrow and, as its mass lies slightly below the
$KK$ threshold, its dominant decay mode is expected to be
$\pi_{\mathsmaller{\!K}}\rightarrow \pi\pi$. The scalar resonance
$f_0(980)$ meets all these expectations~\cite{pdg}.

Let us include now the $\eta$ meson. Taking $h=\eta$ in~\eqref{15}
we predict the characteristics of $\pi_\eta$ resonance:
$m_{\pi_\eta}\approx1010$~MeV, $I_{\pi_\eta}=1$ (it inherits
$I_\pi$), the dominant decay mode $\pi_\eta\rightarrow\eta\pi$ and
it should be broader than $\pi_{\mathsmaller{\!K}}$ because this
mode is not forbidden. The scalar resonance $a_0(980)$ satisfies
these predictions~\cite{pdg}.

Consider a hypothetical $K_\pi$ collisional resonance. The
formation of the coherent state $K_\pi$ is much harder than
$\pi_{\mathsmaller{\!K}}$ because $\pi$ has larger de Broglie
wavelength, i.e. the pion wave packet is larger then the kaon one.
This is also true for the measured mean sizes, $\langle
r_\pi\rangle > \langle r_{\mathsmaller{\!K}}\rangle$~\cite{pdg}.
But one might assume that the probability of $K_\pi$ were non-zero
due to some quantum effects. The Coulomb attraction would favor
then the $K^+_{\pi^-}$ or $K^-_{\pi^+}$ channel, the
$K^\pm_{\pi^0}$ or $K^0_{\pi^\pm}$ are much less plausible. In any
case, the mass of $K_\pi$ would be given by (we make the
replacements $\pi\rightarrow K$ and $h\rightarrow\pi$
in~\eqref{15}),
\begin{equation}
\label{19}
m_{K_{\mathlarger\pi}}^2=\Lambda(m_\pi+m_q+m_s)=\Lambda m_\pi + m_K^2,
\end{equation}
resulting in $m_{K_\pi}\approx710$~MeV. It is tempting to
interpret $K_\pi$ as the unconfirmed scalar resonance $K_0^*(800)$
called also $k$-meson~\cite{pdg}. The Particle Data reports the
following mass of this elusive resonance:
$682\pm29$~MeV~\cite{pdg}. Comparison of~\eqref{19}
with~\eqref{14} shows that, as expected, $K_\pi$ would be a
partner of $\sigma$ in which one of $u$ or $d$ quarks is
substituted by the $s$ quark. The observed isospin $I_k=\frac12$,
however, contradicts the favorable isospin zero predicted by the
assumed mechanism of $K_\pi$ formation. We see thus that the
existence of $k$ meson, if confirmed, is not in conflict with our
general principle for collisional resonances and one should look
for a correct formation mechanism.

We propose the following explanation. Let us take a closer look at
the production of $k$-resonance. The main source of information on
$k$ are decays of $J/\psi$ meson into kaons and pions. The decays
of vector charmonia are always accompanied by an abundant photon
background. In all this "mixture", one can have situations when
photons produce $\pi^0$ meson inside a kaon. The formed coherent
state would then inherit the kaon isospin, i.e. would give rise to
the scalar partners of the pseudoscalar $K^+$, $K^-$, $K^0$, and
$\overline{K}^0$ mesons, with the mass being described by the
relation~\eqref{19}.

Our approach predicts other collisional scalar resonances as well:
$K_\eta$ with mass about $1120$~MeV, $\eta_\eta$ with mass about
$1150$~MeV, almost unfeasible $\eta_\pi$ having mass near
$760$~MeV and a formation mechanism similar to that of $k$ meson,
and resonances with $\eta'$ like the $\pi_{\eta'}$ state of mass
about $1330$~MeV. It is likely very hard to detect these
resonances in the $\pi\pi$, $K\pi$ and $KK$ scattering (in
experiments of direct $\eta\pi$ and $\eta K$ scattering this would
be easier) but they may contribute to the strong background
emerging in these reactions.

Within the framework of our collisional interpretation, the scalar
resonances below and slightly above 1~GeV represent thus two-meson
states. In terms of the quark degrees of freedom, they are
tetraquarks, as is also suggested by many other models and
observations~\cite{pelaez}.

\section{Once more on inclusion of strange quark}

The inclusion of strange quark considered above is not yet a full
extension of our approach to the strange sector. We must switch on
the third quark flavor in the trace of energy-momentum
tensor~\eqref{3}. The $\beta$-function changes slightly in the
gluon part and this numerical change can be neglected in the first
approximation. The main effect comes from the appearance of
operator $m_s\bar{s}s$ in the quark part. The universality of
$\Lambda$ used above is related with the assumption
$\langle\bar{s}s\rangle=\langle\bar{q}q\rangle$ which is
compatible both with the ChPT~\cite{gasser} and  QCD sum
rules~\cite{svz}.

We have already observed an automatic fine-tuning of parameters
$m_\rho\approx E_\rho+E_0$ that enables a dual interpretation of
the axial $h_1$ meson. Our approach contains more interesting
"coincidences". Taking as before $m_s=130$~MeV at the scale 1~GeV,
one can observe the following relations between parameters:
$E_0\approx E_\rho+m_s$ (perhaps, $E_0=E_\rho+m_s+2m_q$) and
$E_1=E_\rho+2m_s$. It is the second relation that leads to
equality of masses of $\varphi$ meson~\eqref{17} and
"materialized" vector underquarkonium $A_1$. The phenomenological
values of $E_0$ and $E_1$ seem to indicate that in quantum world
we cannot build a spectroscopy of excited "pure non-strange"
mesons: If there is available excitation energy
$E_\rho\approx310$~MeV then there is enough energy for creating an
$s\bar{s}$ pair and the strange quarks intervene. The matter looks
as if we added the third quark flavor to~\eqref{4},
$2m_q\rightarrow 2m_q+m_s$, and after that ascribed $m_s$ to a
redefined $E_h$, $E_h+m_s\rightarrow E_h$. The "mass-symmetric"
limit $E_\rho=E_0=E_1$ considered in our discussion of Regge
trajectories means then the limit $m_s=0$. An explanation of
observed "quantization" of $E_0$ and $E_1$ with respect to $m_s$
is challenging\footnote{We would indicate another one
"coincidence": $E_\rho+E_1+m_\pi=2(m_s+E_\rho)+m_\pi=m_\varphi$.
This relation allows of an obvious non-relativistic
interpretation: The $\varphi$ meson is composed of two constituent
strange quarks with mass $m_s+E_\rho$ and binding energy $m_\pi$.
The "free" constituent strange quarks would have the mass
$m_s+E_\rho+m_\pi/2=510$~MeV which is exactly the value usually
extracted from the observed magnetic moment of $\Lambda$ baryon
within a naive quark model~\cite{halzen}.}.

The light non-strange excited mesons have usually some decay modes
with the strange component. Within the standard picture, masses of
these states do not depend on $m_s$ and the $s\bar{s}$ pair pops
up from the vacuum triggering the corresponding decay modes. The
observed relation $E_1\simeq E_0+m_s\simeq E_\rho+2m_s$ may
suggest that, in reality, the strange quark does contribute to
their masses via the contribution of $m_s$ to $E_0$ and $E_1$.

\section{Conclusions}

We have proposed a novel approach to classification of hadrons and
description of hadron spectroscopy. The approach is very simple,
allows of a clear visual interpretation and gives possibility to
estimate hadron masses by a trivial arithmetics on any available
calculator.

In brief, we conjectured an ansatz for hadron masses that should
follow from the trace of energy-momentum tensor in QCD, providing
thereby the relativistic and renormalization invariance of masses.
The Gell-Mann--Oakes--Renner relation for pion mass turns out to
be a particular case of this ansatz. The heavier non-strange
mesons appear via excitation of one of quarks inside pion. A
quark-spectator "feels" a sudden perturbation of energy-momentum
of the excited quark and this affects the hadron mass in the same
manner as a sudden increase of mass of one of quarks. The
excitation is caused either by the gluon exchange  or in the
collisional way in hadron-hadron collisions when a hadron with
smaller wave packet scatters as a whole on one of quarks of
another hadron. For the case of gluon excitations, we developed a
model of "atomic" structure of excited mesons, where the nucleus
is formed by gluon quasiparticles of two kinds --- the
spin-singlet and spin-triplet one. They are not necessarily
colorless and their energies are not renorminvariant. But together
with valent quarks they form a colorless hadron with
renorminvariant mass. The given picture leads to a new systematics
of light mesons above 1~GeV, including the exotic $\pi_1$-states
as an integral part. This systematics does not use the
non-relativistic notion of orbital angular momentum associated
with hadron constituents, or more precisely, it is equal to zero.
The proposed mass counting scheme reproduces various long-known
relations for hadron masses. The "atomic" picture happens to be
smoothly connected with the collisional one near 1~GeV due to a
miraculous fine-tuning of parameters occurring somehow
automatically. The collisional picture explains the whole scalar
sector below 1~GeV in a remarkably simple way --- the sector, that
is the hardest for traditional approaches, turns out to be the
easiest one within the proposed scheme.

In the given paper, the approach was developed for the light
mesons. The presented analysis is partly raw (and perhaps the
constructed scheme is still incomplete) and does not explain all
state-by-state properties of meson spectrum but we consider it as
a possible step forward in a right direction. The extension to the
light baryons is straightforward and is planned for a future work,
together with some theoretical justifications. It seems that many
highly excited $N$ and $\Delta$ baryons can be described as
collisional excitations of the kind $M_{\!B}$, where $M$ is a
meson (typically the $\pi$ meson and resonances which are
abundantly produced in reactions with $\pi$, like $\rho$
($\omega$) and $f_J$ mesons) and $B$ is a baryon (typically the
proton and $\Delta(1232)$). An extension to the heavy quarks is in
progress. In the heavy sector, it is important to understand a
transition from the relativistic to non-relativistic picture. The
interpretation of many unconventional heavy resonances observed
recently~\cite{chen} is especially challenging.

It must be emphasized that the approach proposed in this work is
broader than "just another one model" as it gives a new language
for discussion of hadron resonances, for interpretation of data in
the hadroproduction and formation experiments, and a possible
starting point for construction of essentially new dynamical
models.

%\section*{Acknowledgments}

%The work was partially supported by the RFBR grant 16-02-00348-a.

\end{document}